\documentclass[twocolumn,prb]{revtex4}
\usepackage{amsfonts}
\usepackage[T1]{fontenc}
\usepackage{amsmath,amsbsy,amssymb,graphicx}
\usepackage{times}

\begin{document}

\title{Topological Euler insulators}
\author{Motohiko Ezawa}
\affiliation{Department of Applied Physics, University of Tokyo, Hongo 7-3-1, 113-8656,
Japan}

\begin{abstract}
The Euler number is a new topological number recently debuted in the
topological physics. Unlike the Chern number defined for a band, it is
defined for interbands. We propose a simple model realizing the topological
Euler insulator for the first time. We utilize the fact that the Euler
number in a three-band model in two dimensions is reduced to the Pontryagin
number. A skyrmion structure appears in momentum phase, yielding a
nontrivial Euler number. Topological edge states emerge when the Euler
number is nonzero. We discuss how to realize this model in electric
circuits. We show that topological edge states are well signaled by
impedance resonances.
\end{abstract}

\maketitle

The topological insulator is one of the most prominent concepts found in
this decade\cite{Hasan,Qi}. It is characterized by a nontrivial topological
number defined for bulk. A typical example is the Chern insulator, where
topological phases are indexed by the Chern number. The Chern number is the
integral of the Berry curvature of the wave function describing a single
band. It is characterized by the emergence of chiral edge states in
nanoribbon geometry, which is known as the bulk-edge correspondence.

Recently, the Euler class and the Euler number draw attention in the context
of twisted bilayer graphene\cite{Ahn}, Weyl semimetals\cite{Bouhon} and
quench dynamics\cite{Unal}. The Euler class is a homotopy class of
orthogonal matrices. It is related to "real" eigenfunctions of \ the
Hamiltonian. Furthermore, the Euler number is defined for a set of
interbands. Real eigenfunctions can be protected by PT or C$_{2}$T symmetries\cite{Ahn,Bouhon,Unal}. 
It is contrasted to the Chern number, which is
calculated from "complex" eigenfunctions. Note that\ the Berry connection,
the Berry curvature and the Chern number are zero for real eigenfunctions.

In this paper, we propose a simple model realizing the topological insulator
indexed by the Euler number for the first time, which we name the
topological Euler insulator. We make the use of the fact that the Euler
number is reformulated as the Pontraygin number for the three-band model
with its three eigenfunctions forming an orthonormal basis\cite{Bouhon,Unal}. 
It is interesting that they form a skyrmion structure in momentum space
for a topological insulating phase. Finally, we propose an electric-circuit
implementation of this topological Euler insulator. The band structure is
well observed by impedance resonance.

\textbf{Euler form and Euler number: } We consider a three-band model in two
dimensions with the Hamiltonian $H$. Let $u_{i}\left( \boldsymbol{k}\right) $ be
a real eigenfunction of the Hamiltonian $H$ describing the $i$th band. We
may choose them to form an orthonormal basis,%
\begin{equation}
\langle u_{i}|u_{j}\rangle =\delta _{ij}.  \label{OrthNorm}
\end{equation}%
Then, there holds a relation%
\begin{equation}
\left\vert u_{i}\right\rangle \times \left\vert u_{j}\right\rangle
=\sum_{\ell }\varepsilon _{ij\ell }\left\vert u_{\ell }\right\rangle .
\label{VecN}
\end{equation}%
The Euler form $\mathfrak{E}_{ij}\left( \boldsymbol{k}\right) $ is defined for a
set of two bands,%
\begin{equation}
\mathfrak{E}_{ij}\left( \boldsymbol{k}\right) =\left\langle \nabla u_{i}\left( 
\boldsymbol{k}\right) \right\vert \times \left\vert \nabla u_{j}\left( \boldsymbol{k}\right) \right\rangle ,
\end{equation}%
whose integration over the Brillouin zone yields the Euler number\cite{Bouhon,Unal},%
\begin{equation}
Q_{ij}=\frac{1}{2\pi }\int_{\text{BZ}}\mathfrak{E}_{ij}\left( \boldsymbol{k}\right) dk_{x}dk_{y}.  \label{EulerQ}
\end{equation}%
The eigenfunction $u_{i}\left( \boldsymbol{k}\right) $ has three components,
which we may identify\ with the three-dimensional vector $\boldsymbol{n}_{i}$ as 
$\boldsymbol{n}_{i}=\left\vert u_{i}\right\rangle $, where $i=1,2,3$. Then, the
Pontryagin number is defined by%
\begin{equation}
Q_{i}=\frac{1}{2\pi }\int_{\text{BZ}}\boldsymbol{n}_{i}\cdot \left( \partial
_{k_{x}}\boldsymbol{n}_{i}\times \partial _{k_{y}}\boldsymbol{n}_{i}\right)
dk_{x}dk_{y}  \label{Pont}
\end{equation}%
for each $i$. It is known\cite{Bouhon,Unal} that the Euler number 
(\ref{EulerQ}) is equal to the Pontryagin number (\ref{Pont}) by identifying 
$Q_{\ell }=\frac{1}{2}\sum_{ij}\varepsilon _{\ell ij}Q_{ij}$.

The Berry connection, the Berry curvature and the Chern number are zero in
the present system because eigenfunctions are real.

\textbf{Hamiltonian:} We express the eigenstate\textbf{\ }$n_{3}$\ as%
\begin{equation}
\left\vert u_{3}\right\rangle =\frac{\{m_{1},m_{2},m_{3}\}}{\sqrt{m_{1}^{2}+m_{2}^{2}+m_{3}^{2}}},
\end{equation}%
where $m_{k}$ is a real function of $k_{x}$ and $k_{y}$. The other two
eigenstates obeying Eq.(\ref{OrthNorm}) are given by%
\begin{align}
\left\vert u_{1}\right\rangle &=\frac{\left\{ 0,m_{3},-m_{2}\right\} }{\sqrt{m_{2}^{2}+m_{3}^{2}}},  \label{u1} \\
\left\vert u_{2}\right\rangle &=\frac{\left\{ -\left(
m_{2}^{2}+m_{3}^{2}\right) ,m_{1}m_{2},m_{1}m_{3}\right\} }{\sqrt{m_{2}^{2}+m_{3}^{2}}\sqrt{m_{1}^{2}+m_{2}^{2}+m_{3}^{2}}}.
\end{align}%
We investigate the three-band Hamiltonian given by%
\begin{equation}
H=c_{3}\sqrt{m_{1}^{2}+m_{2}^{2}+m_{3}^{2}}\left\vert u_{3}\right\rangle
\left\langle u_{3}\right\vert +c_{1}\sqrt{m_{2}^{2}+m_{3}^{2}}\left\vert
u_{1}\right\rangle \left\langle u_{1}\right\vert .  \label{HamilA}
\end{equation}%
The eigenvalues read%
\begin{align}
H\left\vert u_{1}\right\rangle &=c_{1}\left( m_{2}^{2}+m_{3}^{2}\right)
\left\vert u_{2}\right\rangle ,\quad H\left\vert u_{2}\right\rangle =0,\qquad
\notag \\
H\left\vert u_{3}\right\rangle &=c_{3}\left(
m_{1}^{2}+m_{2}^{2}+m_{3}^{2}\right) \left\vert u_{3}\right\rangle .
\label{Ene}
\end{align}%
There is a perfect flat bulk band described by $\left\vert
u_{2}\right\rangle $ at zero energy.

\begin{figure}[t]
\centerline{\includegraphics[width=0.48\textwidth]{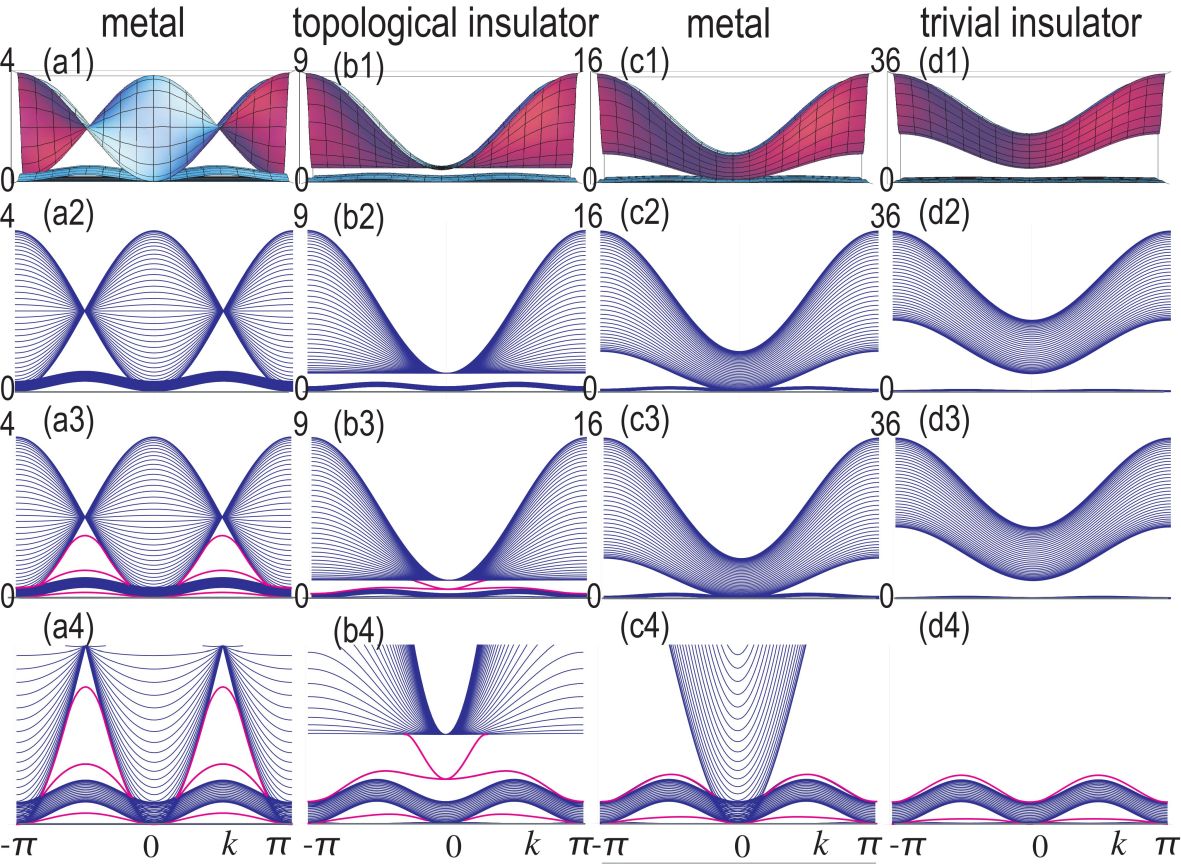}}
\caption{Band structures of a nanoribbon with 40-site width. The horizontal
axis is the momentum $k$, and the vertical axis is the energy in unit of $t$. 
(a) Metal phase with $\protect\mu =0 $, (b) topological insulator phase
with $\protect\mu =t$, (c) metal phase with $\protect\mu =2t$, and (d)
trivial insulator phase with $\protect\mu =4t $. We have set $\protect\lambda =t$. 
(*1) Bird's eye view of the bulk band. (*2) Projection of the
bulk bands. (*3) Band structure of a nanoribbon. The edge states are marked
in red. (*4) Enlarged band structure of (*3) near the zero energy. We have
set $c_{3}=1$ and $c_{1}=0.25.$}
\label{FigRibbon}
\end{figure}

In this work, for definiteness, we study such an explicit model that%
\begin{align}
m_{1}& =t\left( \cos k_{x}+\cos k_{y}\right) -\mu ,  \notag \\
m_{2}& =\lambda \sin k_{x},\quad m_{3}=\lambda \sin k_{y}.  \label{FormM}
\end{align}%
The band gap is calculated from (\ref{Ene}) as 
\begin{equation}
\Delta =\text{min}\left[ \left\vert \pm 2t-\mu \right\vert ,\left\vert \mu
\right\vert \right]
\end{equation}%
at $\left( k_{x},k_{y}\right) =\left( 0,0\right) $, $\left( 0,\pi \right) $, 
$\left( \pi ,0\right) $ or $\left( \pi ,\pi \right) $. The band closes at $\mu =\pm 2t$ as in Fig.\ref{FigRibbon}.

\begin{figure}[t]
\centerline{\includegraphics[width=0.48\textwidth]{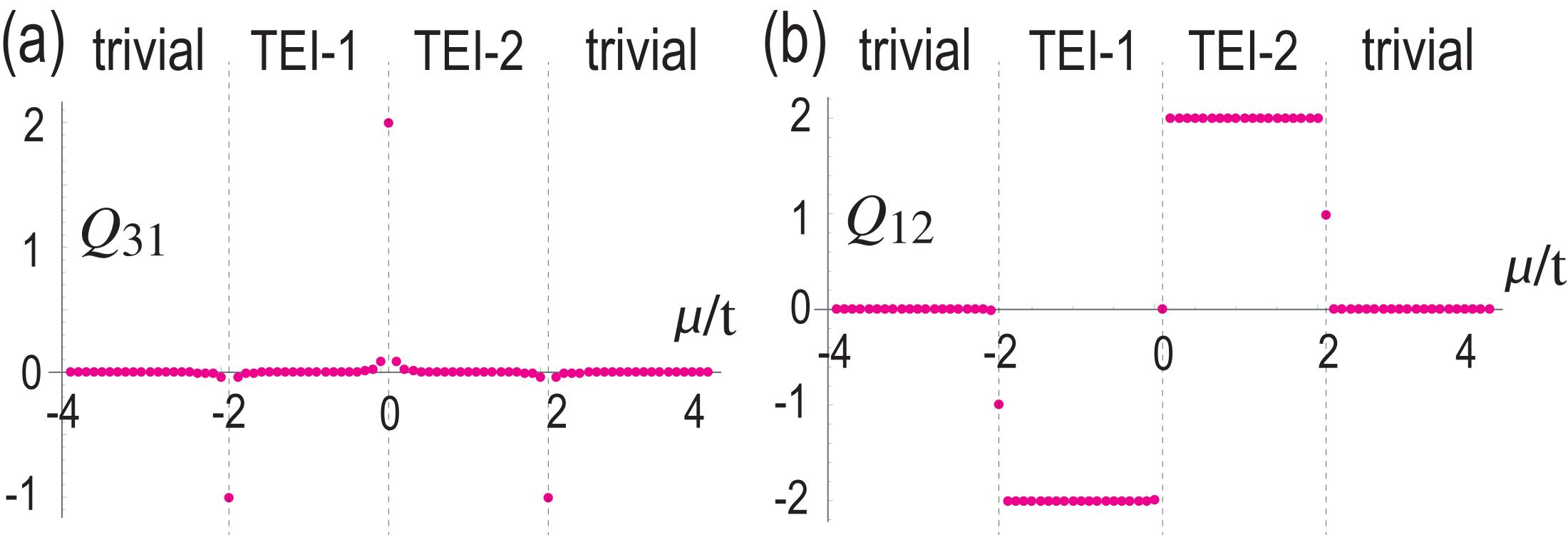}}
\caption{Euler number as a function of $\protect\mu /t$ for (a) $Q_{31}$ and
(b) $Q_{12}$. We note that $Q_{23}=0$ for all $\protect\mu $. There are two
topological-Euler-insulator phases denoted by TEI-1 and TEI-2. The
horizontal axis is $\protect\mu /t$.}
\label{FigSkyrmion}
\end{figure}

\begin{figure}[t]
\centerline{\includegraphics[width=0.48\textwidth]{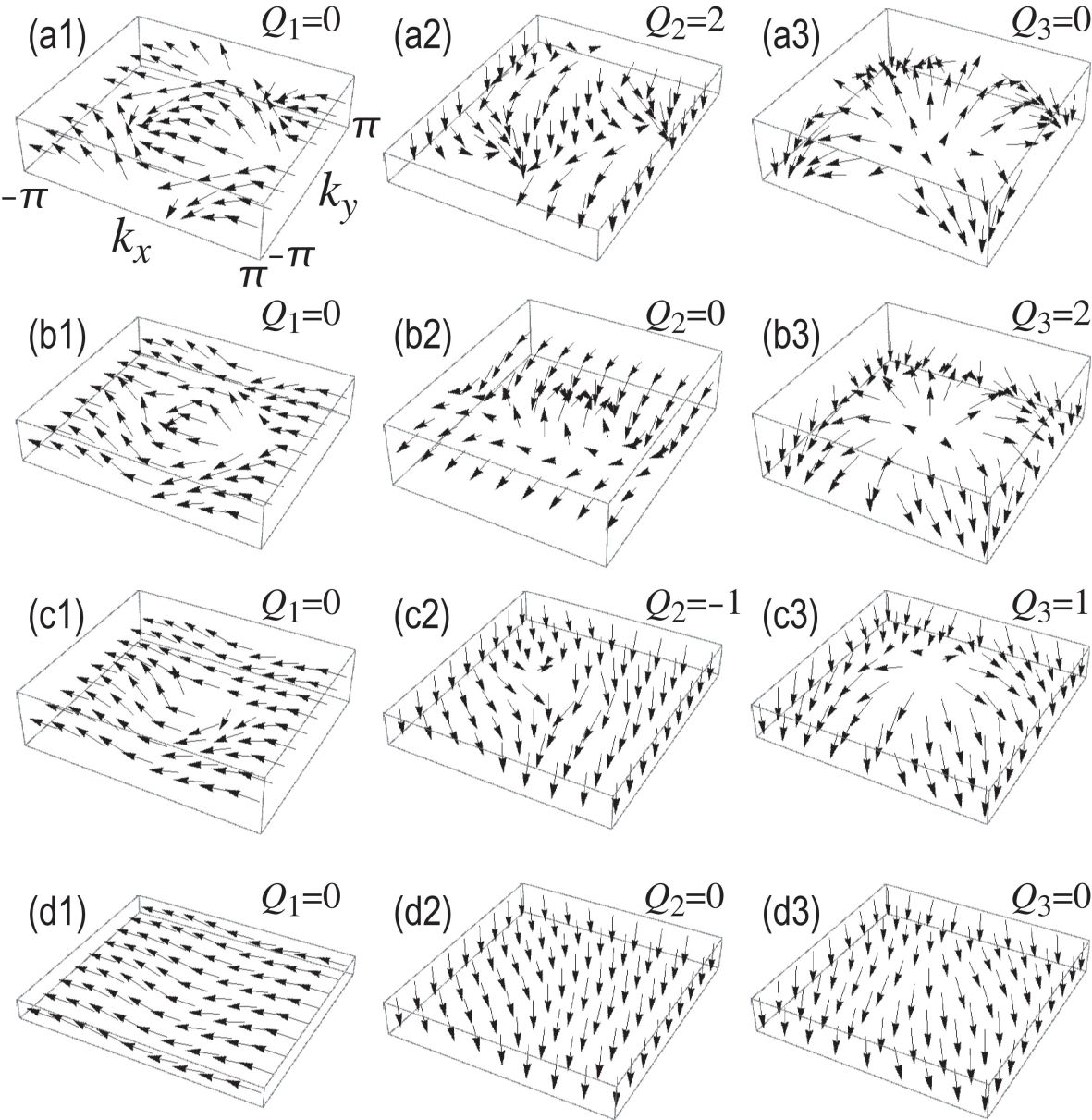}}
\caption{Spin textures in the $k_x$-$k_y$ plane for (*1) $\boldsymbol{n}_{1}$,
(*2) $\boldsymbol{n}_{2}$ and (*3) $\boldsymbol{n}_{3}$, yielding the Pontryagin
numbers $Q_1$, $Q_2$ and $Q_3$. (a*) Metal phase with $\protect\mu =0$, (b*)
topological insulator phase with $\protect\mu =t$, (c*) metal phase with 
$\protect\mu =2t$, and (d*) trivial insulator phase with $\protect\mu =4t$.
The corresponding Euler numbers are shown in Fig.\protect\ref{FigSkyrmion}.}
\label{FigSpin}
\end{figure}

\begin{figure}[t]
\centerline{\includegraphics[width=0.48\textwidth]{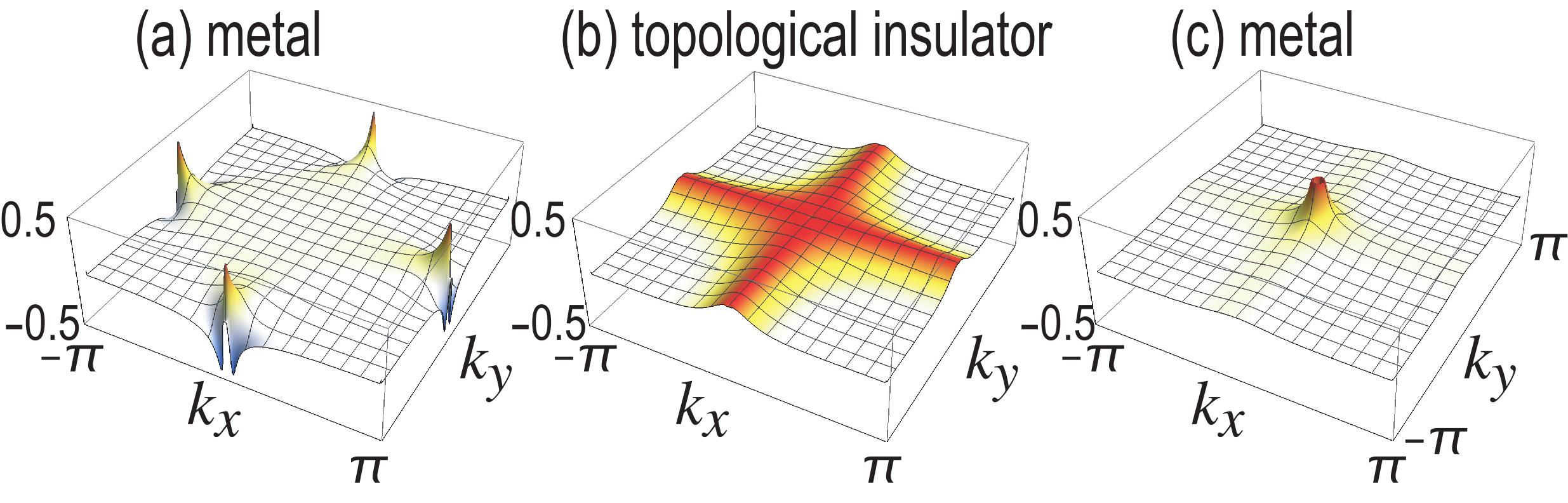}}
\caption{Local Pontraygin density in the $k_x$-$k_y$ plane for (a) metal
phase with $\protect\mu =0$, (b) topological insulator phase with $\protect\mu =t$, 
and (c) metal phase with $\protect\mu =2t$. }
\label{FigSkDOS}
\end{figure}

\textbf{Topological number:} We study the Euler number (\ref{EulerQ}) or
equivalently the Pontryagin number (\ref{Pont}). They are related as 
$Q_{23}=Q_{1}$, $Q_{31}=Q_{2}$ and $Q_{12}=Q_{3}$. We have numerically
integrated the Pontryagin density, whose results are shown in Fig.\ref{FigSkyrmion}. 
The Pontryagin number is graphically understood by plotting
the direction of the vectors $\boldsymbol{n}_{i}$, as shown in Fig.\ref{FigSpin}. 
It is nonzero only when the texture forms a skyrmion. It may change its
value when the bulk band gap closes. It indicates that our system is a kind
of topological insulators, which we call topological Euler insulators.

\begin{figure}[t]
\centerline{\includegraphics[width=0.48\textwidth]{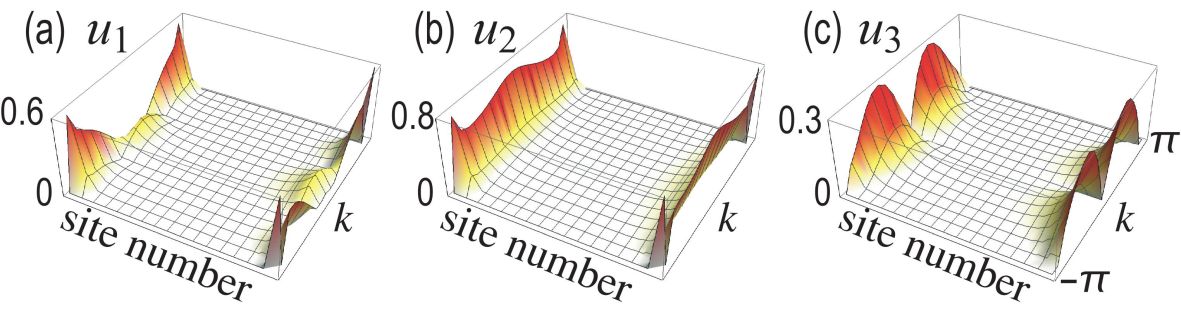}}
\caption{Absolute value of the eigenfunction for localized edge states in a
nanoribbon with 40-site width. (a), (b) and (c) are for the bands 
$|u_{1}\rangle $, $|u_{2}\rangle $ and $|u_{3}\rangle $. They are well
localized at the edges. }
\label{FigEdge}
\end{figure}

The Pontryagin number reads as follows. First we find $Q_{1}=0$. It is
understood by the fact that the configuration (\ref{u1}) is restricted
within the $yz$ plane. Next, we find $Q_{2}=2$ for $\mu =0$ and $Q_{2}=-1$
for $\mu =\pm t$, and $Q_{2}=0$ otherwise, as shown in Fig.\ref{FigSkyrmion}(a).

Finally, we find $Q_{3}=2$ for $0<\mu <2t$, $Q_{3}=-2$ for $-2t<\mu <0$, 
$Q_{3}=\pm 1$ for $\mu =\pm 2t$, and $Q_{3}=0$ otherwise, as shown in Fig.\ref{FigSkyrmion}(b). 
The local Pontryagin density $Q_{3}\left(
k_{x},k_{y}\right) \equiv \boldsymbol{n}_{3}\cdot \left( \partial _{k_{x}}
\boldsymbol{n}_{3}\times \partial _{k_{y}}\boldsymbol{n}_{3}\right) $ is given in
Fig.\ref{FigSkDOS}. It is interesting to note that the same topological
phase diagram is produced by the following two-band model,%
\begin{equation}
H_{2\text{band}}=\sum_{j=1,2,3}m_{j}\sigma _{j},
\end{equation}%
where $m_{j}$ are given by (\ref{FormM}) and $\sigma _{j}$ are the Pauli
matrices. This describes a Chern insulator for $0<\left\vert \mu \right\vert
<2t$.

In conclusion, the system is topological for $0<\left\vert \mu \right\vert
<2t$. The corresponding edge states emerge in the topological phase, as we
now see.

\textbf{Bulk-edge correspondence:} We show the band structures of a
nanoribbon in Fig.\ref{FigRibbon}. They coincide with those of bulk
Hamiltonian except for the edge states marked in magenta. The eigenfunction
is well localized at the edges as shown in Fig.\ref{FigEdge}.

It is possible to study the edge states analytically. First, we study them
at $k_{y}=\pi $, where the eigenequation is obtained from Eq.(\ref{HamilA})
as%
\begin{equation}
\mu \left( 
\begin{array}{ccc}
0 & i\lambda \partial _{x} & 0 \\ 
i\lambda \partial _{x} & \mu & 0 \\ 
0 & 0 & 0%
\end{array}%
\right) \left( 
\begin{array}{c}
\left\vert u_{1}\right\rangle \\ 
\left\vert u_{2}\right\rangle \\ 
\left\vert u_{3}\right\rangle%
\end{array}%
\right) =E\left( 
\begin{array}{c}
\left\vert u_{1}\right\rangle \\ 
\left\vert u_{2}\right\rangle \\ 
\left\vert u_{3}\right\rangle%
\end{array}%
\right)  \label{EqA}
\end{equation}%
in the first order of $k_{x}$. We find $\left\vert u_{3}\right\rangle =0$,
as agrees with the numerical result in Fig.\ref{FigEdge}(c). By assuming
exponentially damping eigenfunctions for $\left\vert u_{1}\right\rangle $
and $\left\vert u_{2}\right\rangle $,
\begin{equation}
\left\vert u_{1}\right\rangle =c_{a}e^{\kappa x},\qquad \left\vert
u_{2}\right\rangle =\pm ic_{a}e^{\kappa x},
\end{equation}%
we obtain from Eq.(\ref{EqA}) that $\pm \mu \lambda \kappa =E$ and $-\mu
\lambda \kappa \pm \mu ^{2}=\pm E$. We solve the energy $E$ and the
penetration depth $\kappa $ as%
\begin{equation}
E=\frac{\mu ^{2}}{2},\quad \kappa =\pm \frac{\mu }{2\lambda },
\end{equation}%
where $\kappa =\pm \mu /2\lambda $\ describe the edge states localized at
the right edge and the left edge, respectively. They agree well with the
numerical results in Fig.\ref{FigRibbon}(b4) at $k=\pi $.

Next we study the edge states at $k_{y}=0$, where we have 
\begin{equation}
\left( 2t-\mu \right) \left( 
\begin{array}{ccc}
0 & -i\lambda \partial _{x} & 0 \\ 
-i\lambda \partial _{x} & 2t-\mu & 0 \\ 
0 & 0 & 0%
\end{array}%
\right) \left( 
\begin{array}{c}
\left\vert u_{1}\right\rangle \\ 
\left\vert u_{2}\right\rangle \\ 
\left\vert u_{3}\right\rangle%
\end{array}%
\right) =E\left( 
\begin{array}{c}
\left\vert u_{1}\right\rangle \\ 
\left\vert u_{2}\right\rangle \\ 
\left\vert u_{3}\right\rangle%
\end{array}%
\right)
\end{equation}%
in the first order of $k_{x}$. We have $\left\vert u_{3}\right\rangle =0$,
as agrees with the numerical result in Fig.\ref{FigEdge}(c). We have
solutions%
\begin{equation}
E=\frac{\left( 2t-\mu \right) ^{2}}{2},\quad \kappa =\mp \frac{2t-\mu }{2\lambda }.
\end{equation}%
These solutions show localized edge states at nonzero energies. They agree
well with the numerical results in Fig.\ref{FigRibbon}(b4) at $k=0$.

\textbf{Electric-circuit simulations: } Electric circuits are characterized
by the Kirchhoff current law. By making the Fourier transformation with
respect to time, the Kirchhoff current law is expressed as%
\begin{equation}
I_{a}\left( \omega \right) =\sum_{b}J_{ab}\left( \omega \right) V_{b}\left(
\omega \right) ,  \label{CircuLap}
\end{equation}%
where $I_{a}$ is the current between node $a$ and the ground, while $V_{b}$
is the voltage at node $b$. The matrix $J_{ab}\left( \omega \right) $ is
called the circuit Laplacian. Once the circuit Laplacian is given, we can
uniquely setup the corresponding electric circuit. By equating it with the
Hamiltonian $H$ as\cite{TECNature,ComPhys} 
\begin{equation}
J_{ab}\left( \omega \right) =i\omega H_{ab}\left( \omega \right) ,
\label{JH}
\end{equation}%
it is possible to simulate various topological phases of the Hamiltonian by
electric circuits\cite%
{TECNature,ComPhys,Hel,Lu,Research,Zhao,YLi,EzawaTEC,EzawaLCR,EzawaSkin,Garcia,Hofmann,EzawaMajo,Tjunc,Lee}. 
The relations between the parameters in the Hamiltonian and in the
electric circuit are determined by this formula.

In order to derive the circuit Laplacian, we explicitly write down the
components of the Hamiltonian (\ref{HamilA}), 
\begin{align}
H_{11}=& c_{3}[\frac{t^{2}}{2}\left( \cos 2k_{x}+\cos 2k_{y}+4\cos k_{x}\cos
k_{y}+2\right)  \notag \\
& -2\mu t\left( \cos k_{x}+\cos k_{y}\right) +\mu ^{2}], \\
H_{22}=& \frac{\lambda ^{2}}{2}[c_{3}\left( 1-\cos 2k_{x}\right)
+c_{1}\left( 1-\cos 2k_{y}\right) ], \\
H_{33}=& \frac{\lambda ^{2}}{2}[c_{1}\left( 1-\cos 2k_{y}\right)
+c_{3}\left( 1-\cos 2k_{y}\right) ],
\end{align}%
and%
\begin{align}
H_{12}& =c_{3}[t\lambda \sin k_{x}\left( \cos k_{x}+\cos k_{y}\right) -\mu
\lambda \sin k_{x}], \\
H_{13}& =c_{3}[t\lambda \sin k_{y}\left( \cos k_{x}+\cos k_{y}\right) -\mu
\lambda \sin k_{y}]. \\
H_{23}& =\left( c_{3}-c_{1}\right) \lambda ^{2}\sin k_{x}\sin k_{y}.
\end{align}%
Here, we make a convention that $t$, $\mu $ and $\lambda $ are
dimensionless, and hence that $c_{1}$ and $c_{2}$ have the dimension of
energy.

The circuit Laplacian is constructed as follows. To simulate the positive
and negative hoppings in the Hamiltonian, we replace them with the
capacitance $i\omega C$ and the inductance $1/i\omega L$, respectively. We
note that $\sin k=(e^{ik}-e^{-ik})/2i$ represents an imaginary hopping in
the tight-bind model. The imaginary hopping is realized by an operational
amplifier\cite{Hofmann}.

We thus make the following replacements with respect to hoppings in the
Hamiltonian to derive the circuit Laplacian: (i) $+X\rightarrow i\omega
C_{X} $ for $X=t^{2},\mu ^{2},\lambda ^{2}$, where $C_{X}$ represents the
capacitance whose value is $X$ [pF]. (ii) $-X\rightarrow 1/i\omega L_{X}$
for $X=\mu t,\lambda ^{2}$, where $L_{X}$ represents the inductance whose
value is $X$ [$\mu $H]. (iii) $\pm iX\rightarrow \pm 1/R_{X}$ for 
$X=t\lambda ,\mu \lambda $, where $R_{X}$ represents the resistance whose
value is $X$ [k$\Omega $].

Consequently, we obtain%
\begin{align}
J_{11}=& \frac{i\omega c_{3}C_{t^{2}}}{2}\left( \cos 2k_{x}+\cos
2k_{y}+4\cos k_{x}\cos k_{y}+2\right)  \notag \\
& +2\frac{1}{i\omega L_{\mu t}}\left( \cos k_{x}+\cos k_{y}\right) +i\omega
C_{\mu ^{2}}, \\
J_{22}=& \frac{c_{3}}{2}\left( i\omega C_{\lambda ^{2}}+\frac{1}{i\omega
L_{\lambda ^{2}}}\cos 2k_{x}\right)  \notag \\
& +\frac{c_{1}}{2}\left( i\omega C_{\lambda ^{2}}+\frac{1}{i\omega
L_{\lambda ^{2}}}\cos 2k_{y}\right) , \\
J_{33}=& \frac{c_{1}}{2}\left( i\omega C_{\lambda ^{2}}+\frac{1}{i\omega
L_{\lambda ^{2}}}\cos 2k_{x}\right)  \notag \\
& +\frac{c_{3}}{2}\left( i\omega C_{\lambda ^{2}}+\frac{1}{i\omega
L_{\lambda ^{2}}}\cos 2k_{y}\right) ,
\end{align}%
and%
\begin{align}
J_{12}& =\frac{c_{3}}{iR_{t\lambda }}\sin k_{x}\left( \cos k_{x}+\cos
k_{y}\right) -\frac{c_{3}}{R_{\lambda \mu }}\sin k_{x}, \\
J_{13}& =\frac{c_{3}}{iR_{t\lambda }}\sin k_{x}\left( \cos k_{x}+\cos
k_{y}\right) -\frac{c_{3}}{R_{\lambda \mu }}\sin k_{x}, \\
J_{23}& =i\omega \left( c_{3}-c_{1}\right) C_{\lambda ^{2}}\sin k_{x}\sin
k_{y}.
\end{align}%
Furthermore, in order to realize impedance resonances, we add a circuit
corresponding to\cite{TECNature,ComPhys,EzawaTEC,EzawaMajo,Tjunc}%
\begin{equation}
\Delta J_{11}=\Delta J_{22}=\Delta J_{33}=i\omega C_{0}+\frac{1}{i\omega
L_{0}},
\end{equation}%
which vanish at the critical frequency $\omega _{0}=1/\sqrt{L_{0}C_{0}}$.

After the diagonalization, the circuit Laplacian yields 
\begin{equation}
J_{n}\left( \omega \right) =i\omega C_{0}+\frac{1}{i\omega L_{0}}-i\omega
\varepsilon _{n}\left( \omega \right) ,
\end{equation}%
where $\varepsilon _{n}$ is the eigenvalue of the circuit Laplacian. Solving 
$J_{n}\left( \omega \right) =0$, we obtain 
\begin{equation}
\omega _{\text{R}}(\varepsilon _{n})=\sqrt{L_{0}/(C_{0}-\varepsilon _{n})},
\label{ResonFrequ}
\end{equation}%
which corresponds to the impedance resonance frequency.

\begin{figure}[t]
\centerline{\includegraphics[width=0.48\textwidth]{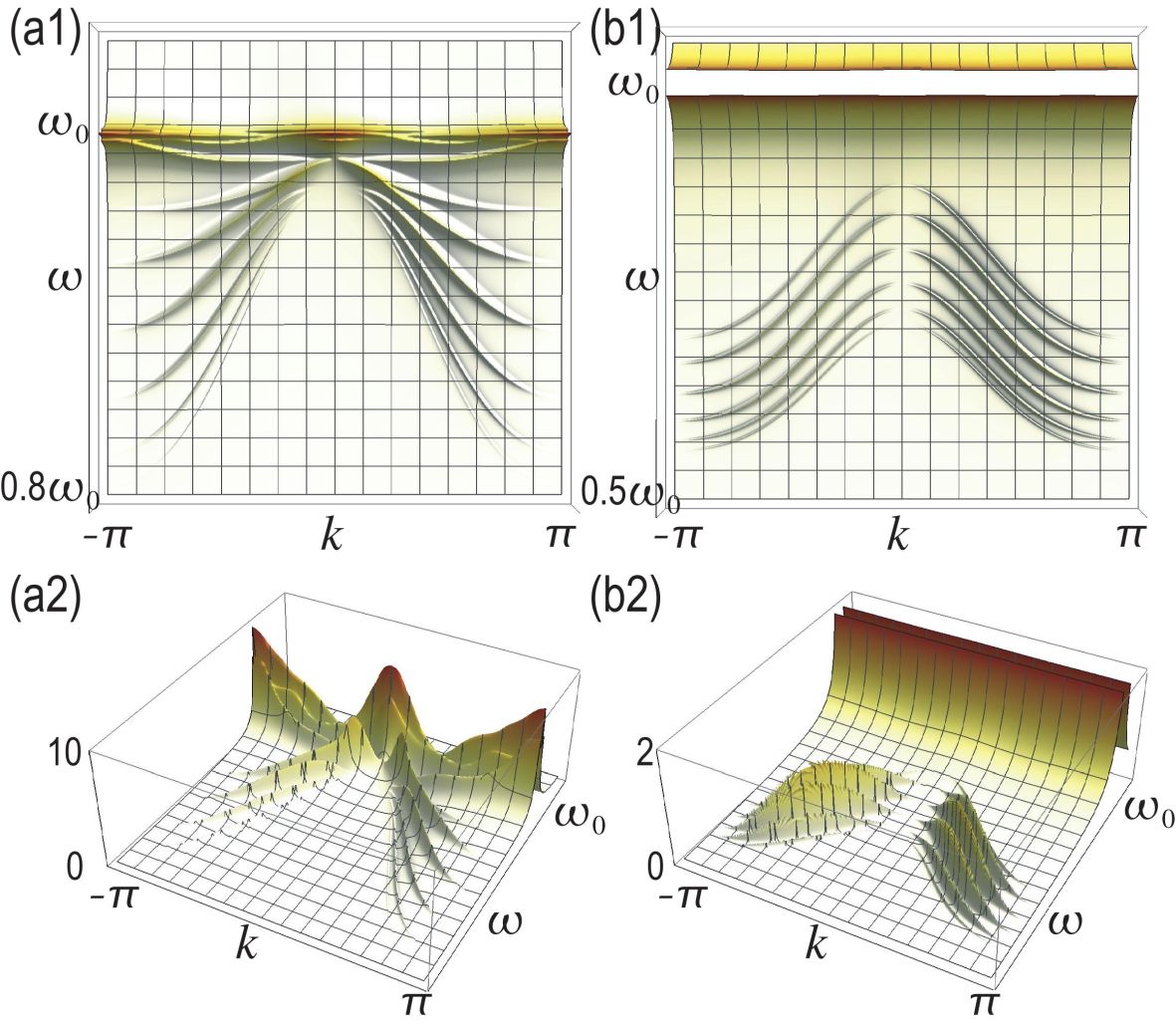}}
\caption{Impedance in the $k$-$\protect\omega $ plane. (a*) Topological
phase with $\protect\mu =-t$ and (b*) trivial phase with $\protect\mu =-4t$.
(*1) Top view and (*2) bird's eye's view. We have set $c_{3}=1$ and $c_{1}=0.25$. 
The vertical axis is the impedance in unit of k$\Omega $.}
\label{FigImpe}
\end{figure}

Topological edge states are observed by impedance resonances, where the
impedance between nodes $a$ and $b$ is given by\cite{Hel} $Z_{ab}\equiv
V_{a}/I_{a}=G_{ab}$, where $G$ is the Green function defined by the inverse
of the circuit Laplacian $J$, $G\equiv J^{-1}$. The momentum-dependent
impedance is an experimentally detectable quantity\cite{Hel,Lee} by using a
Fourier transformation along the nanoribbon direction $y$,%
\begin{equation}
Z_{\alpha \beta }\left( x,k_{y},\omega \right) =\sum_{\rho }Z_{\alpha \beta
}\left( x_{\rho },y_{\rho },\omega \right) \exp \left[ -iy_{\rho }k_{y}%
\right] ,
\end{equation}%
where $\left( x_{\rho },y_{\rho }\right) $ is the Bravais vector. We take $\alpha =\beta $ at edge sites.

We show an impedance as a function of the momentum $k$ and the frequency $\omega $ in Fig.\ref{FigImpe}. 
Topological edge states are clearly observed
in the topological phase. Note that the top and the bottom are reversed
between the band structure and the impedance resonance, as indicated in Eq.(\ref{ResonFrequ}), 
where $\varepsilon _{n}$ represents the band structure
while $\omega _{\text{R}}$ the impedance resonance.

In this work we have proposed topological Euler insulators, which are
characterized by nontrivial Euler numbers. Their band structure including
the edge states is well observed by measuring the impedance of the
corresponding electric circuit.

The author is very much grateful to N. Nagaosa for helpful discussions on
the subject. This work is supported by the Grants-in-Aid for Scientific
Research from MEXT KAKENHI (Grants No. JP17K05490 and No. JP18H03676). This
work is also supported by CREST, JST (JPMJCR16F1 and JPMJCR20T2).


\begin{thebibliography}{99}
\bibitem{Hasan} M. Z. Hasan and C. L. Kane, Rev. Mod. Phys. \textbf{82},
3045 (2010).

\bibitem{Qi} X.-L. Qi and S.-C. Zhang, Rev. Mod. Phys. \textbf{83}, 1057
(2011).

\bibitem{Ahn} J. Ahn, S. Park and B. J. Yang, Phys. Rev. X \textbf{9}, 021013 (2019).

\bibitem{Bouhon} A. Bouhon, Q. Wu, R.-J. Slager, H. Weng, O. V. Yazyev and
T. Bzdusek, Nature Physics (2020).

\bibitem{Unal} F. Nur Unal, A. Bouhon and R.-J. Slager, Phys. Rev. Lett.
\textbf{125}, 053601 (2020).

\bibitem{TECNature} S. Imhof, C. Berger, F. Bayer, J. Brehm, L. Molenkamp,
T. Kiessling, F. Schindler, C. H. Lee, M. Greiter, T. Neupert, R. Thomale,
Nat. Phys. \textbf{14}, 925 (2018).

\bibitem{ComPhys} C. H. Lee , S. Imhof, C. Berger, F. Bayer, J. Brehm, L. W.
Molenkamp, T. Kiessling and R. Thomale, Communications Physics, \textbf{1},
39 (2018).

\bibitem{Hel} T. Helbig, T. Hofmann, C. H. Lee, R. Thomale, S. Imhof, L. W.
Molenkamp and T. Kiessling, Phys. Rev. B \textbf{99}, 161114 (2019).

\bibitem{Lu} Y. Lu, N. Jia, L. Su, C. Owens, G. Juzeliunas, D. I. Schuster
and J. Simon, Phys. Rev. B \textbf{99}, 020302 (2019).

\bibitem{Research} K. Luo, R. Yu and H. Weng, Research (2018), ID 6793752.

\bibitem{Zhao} E. Zhao, Ann. Phys. \textbf{399}, 289 (2018).

\bibitem{YLi} Y. Li, Y. Sun, W. Zhu, Z. Guo, J. Jiang, T. Kariyado, H. Chen
and X. Hu, Nat. Com. \textbf{9}, 4598 (2018).

\bibitem{EzawaTEC} M. Ezawa, Phys. Rev. B \textbf{98}, 201402(R) (2018).

\bibitem{Garcia} M. Serra-Garcia, R. Susstrunk and S. D. Huber, Phys. Rev. B 
\textbf{99}, 020304 (2019).

\bibitem{Hofmann} T. Hofmann, T. Helbig, C. H. Lee, M. Greiter, R. Thomale,
Phys. Rev. Lett. \textbf{122}, 247702 (2019).

\bibitem{EzawaMajo} M. Ezawa, Phys. Rev. B \textbf{100}, 045407 (2019)

\bibitem{EzawaLCR} M. Ezawa, Phys. Rev. B \textbf{99}, 201411(R) (2019).

\bibitem{EzawaSkin} M. Ezawa, Phys. Rev. B \textbf{99}, 121411(R) (2019).

\bibitem{Tjunc} M. Ezawa, Phys. Rev. B \textbf{102}, 075424 (2020).

\bibitem{Lee} C. H. Lee, T. Hofmann, T. Helbig, Y. Liu, X. Zhang, M. Greiter
and R. Thomale, Nature Communications, \textbf{11}, 4385 (2020).
\end{thebibliography}
\end{document}